\documentclass[prl,twocolumn]{revtex4}

\usepackage{graphicx}

\newcommand\ba{{\mathbf{a}}}
\newcommand\bb{{\mathbf{b}}}
\newcommand\bff{{\mathbf{f}}}
\newcommand\bx{{\mathbf{x}}}
\newcommand\bxd{\dot{\mathbf{x}}}
\newcommand\bX{{\mathbf{X}}}
\newcommand\bXd{\dot{\mathbf{X}}}
\newcommand\bzero{{\mathbf{0}}}
\newcommand\pa{\partial}
\newcommand\beq{\begin{equation}}
\newcommand\eeq{\end{equation}}
\newcommand\bea{\begin{eqnarray}}
\newcommand\eea{\end{eqnarray}}
\newcommand\cd{\cdot}
\newcommand\al{\alpha}
\newcommand\ga{\gamma}
\newcommand\de{\delta}

\newcommand\si{\sigma}
\newcommand\ta{\tau}

\begin{document}

\title{Collisions of strings with Y junctions}

\author{E.~J.~Copeland}
\email{ed.copeland@nottingham.ac.uk} \affiliation{School of
Physics and Astronomy, University of Nottingham, University Park,
Nottingham NG7 2RD, United Kingdom}
\author{T.~W.~B.~Kibble}
\email{kibble@imperial.ac.uk}
\affiliation{Blackett Laboratory, Imperial College, London SW7 2AZ, United Kingdom}
\author{D.~A.~Steer}
\email{steer@apc.univ-paris7.fr} \affiliation{APC, UMR 7164, 11
place Marcelin Berthelot, 75005 Paris, France, and \\ LPT,
Universit{\' e} de Paris-Sud, Bat. 210, 91405  Orsay  Cedex,
France.}

\date{\today}

\begin{abstract}
We study the dynamics of Nambu--Goto strings with junctions at
which three strings meet.  In particular, we exhibit one simple
exact solution and examine the process of intercommuting of two
straight strings, in which they exchange partners but become
joined by a third string.  We show that there are important
kinematical constraints on this process.  The exchange cannot
occur if the strings meet with very large relative velocity.  This
may have important implications for the evolution of cosmic superstring networks and non-abelian string networks.
\end{abstract}


\maketitle

There has been renewed interest in cosmic strings, both because of
tentative observational evidence
\cite{Sazhin:2003cp,Sazhin:2004fv} and because they appear to
arise naturally in scenarios based on string theory
\cite{Polchinski:2004hb,Kibble:2004hq,Davis:2005dd}, as well as in
field theories \cite{Hindmarsh:1994re,Vilenkin:1994}.   Although
the recent close inspection by the Hubble Space Telescope of the
area of interest in \cite{Sazhin:2003cp,Sazhin:2004fv} appears to
indicate that  no string is present \cite{Sazhin:2006fe}, the
possibility remains that strings may be found through other types
of observation in the Universe. Moreover, in the scenarios based
on string theory, several different kinds of cosmic strings may
appear, in particular F- and D-strings and $(p,q)$ composites,
formed of $p$ F-strings and $q$ D-strings \cite{Copeland:2003bj}.
In such cases, junctions may form at which three different strings
meet.  Such junctions can also appear in networks of `non-Abelian
strings', for which the fundamental group $\pi_1(\mathcal{M})$ of
the manifold of degenerate vacua, which classifies the strings, is
non-Abelian \cite{Alford:1992yx,Hashimoto:2005hi}.  Several papers
have considered the evolution of networks of strings with
junctions
\cite{Vachaspati:1986cc,Spergel:1996ai,Tye:2005fn,Copeland:2005cy},
in particular the question of whether such a network would evolve
to a scaling regime as expected for an ordinary cosmic-string
network \cite{Austin:1994dz}.

In this paper, we study the dynamics of three-string junctions in a local-string network, that is to say one where the individual strings have no long-range interactions and are well described by the Nambu--Goto action.  Our approach is similar to the one adopted by 't Hooft in Ref.~\cite{'tHooft:2004he}, in which he represented baryons as pieces of open string connected at one common point.  However, our method differs from his in significant ways.  In particular, our treatment applies to strings with different tensions, and we use a temporal world-sheet coordinate equal to the global time.  For ordinary cosmic strings, the existence of exact solutions for oscillating string loops \cite{Kibble:1982cb,Burden:1985md} was important in analyzing the likely behavior of loops in general, and some exact solutions are also known for open strings with junctions.   
Here we give one very simple example of an exact solution, but  our main focus is on the question of what happens when two strings cross.

When two ordinary cosmic strings intersect they normally `intercommute', or exchange partners \cite{Shellard:1987bv,Polchinski:1988cn,Copeland:1986ng,Shellard:1988zx,Matzner:1988}.  But for the strings we are considering this is generally impossible.  What we expect instead is that the strings will become joined by a third string.  The dynamical problem of finding the intercommuting probability for junction-forming strings has been discussed by several authors \cite{Jackson:2004zg,Hanany:2005bc}. Our study, of the interaction of a pair of straight strings, is complementary.  We will show that there are important \emph{kinematical} constraints implying that such intercommuting is impossible for strings that meet with very high relative velocity.  The limit depends on the angle at which the strings meet and on the ratios of the string tensions.  (This does not appear to be related to the bound found in simulations in Ref.~\cite{Hanany:2005bc}, which has a quite different dependence on the angle.)  Although colliding ($p,q$) strings generally have different tensions, we consider explicitly here only the case in which the two initial strings have equal tension --- though the third string that joins them may have a different tension --- because the inherent symmetry of the problem makes this relatively easy to solve.  However, similar limits apply more generally, as we will show in a later publication.

We first review the equations of motion of strings with junctions. We use the standard conformal gauge conditions, in which the temporal world-sheet coordinate is identified with the time, $\ta=t$, and the spatial coordinates $\bx(\si,t)$ satisfy the gauge conditions
 \beq \bxd\cd\bx'=0, \qquad \bxd^2+\bx'{}^2=1, \label{gc}\eeq
where $\bxd=\pa_t\bx$ and $\bx'=\pa_\si\bx$.

Let us consider a junction of three strings, with coordinates
$\bx_j(\si,t)$ and string tensions $\mu_j$, $(j=1,2,3)$.  (We take
$\si$ to be increasing towards the vertex on all three strings.)
The position of the vertex is denoted by $\bX$ and the values of
the spatial world-sheet coordinates $\si$ there by $s_j(t)$.  Then
the action may be written
 \bea S&=&-\sum_j\mu_j\int dt\int d\si\,\theta(s_j(t)-\si)\sqrt{\bx'_j{}^2(1-\bxd_j^2)}
 \nonumber\\
 &&\qquad+\sum_j\int dt\,\bff_j(t)\cd[\bx_j(s_j(t),t)-\bX(t)],
 \label{action} \eea
where the $\bff_j$ are Lagrange multipliers.

Varying $\bx_j$ and using (\ref{gc}) yields the usual equation of motion
 \beq \ddot\bx_j-\bx_j''=\bzero,\label{em} \eeq
but there are also boundary terms proportional to $\de(s_j(t)-\si)$ which give
 \beq \mu_j(\bx'_j+\dot s_j\bxd_j)=\bff_j,\label{f} \eeq
where the functions are evaluated at $(s_j(t),t)$.  Varying the Lagrange multipliers $\bff_j$ of course provides the boundary conditions
 \beq \bx_j(s_j(t),t)=\bX(t), \label{X}\eeq
while varying $\bX$ provides the constraint
 \beq \sum_j\bff_j=\bzero.\label{sum} \eeq
Finally, varying $s_j$ and using (\ref{gc}) again, we get
 \beq \bff_j\cd\bx'_j=\mu_j\bx'_j{}^2, \eeq
which is not an independent equation but an immediate consequence of (\ref{f}).

The general solution of (\ref{em}) is of course
 \beq \bx_j(\si,t)=\frac{1}{2}[\ba_j(\si+t)+\bb_j(\si-t)], \eeq
where to satisfy the gauge condition (\ref{gc}) we require
 \beq \ba'_j{}^2=\bb'_j{}^2=1. \eeq
Thus (\ref{X}) becomes
 \beq \ba_j(s_j+t)+\bb_j(s_j-t)=2\bX(t).\label{abX} \eeq
In addition, from (\ref{f}) and (\ref{sum}) we have
 \beq \sum_j\mu_j[(1+\dot s_j)\ba'_j+(1-\dot s_j)\bb'_j]=\bzero.
 \label{sumab}\eeq

The initial conditions for $\bx_j$ and $\bxd_j$ at $t=0$ serve to fix the functions $\ba_j(\si)$ and $\bb_j(\si)$ for $\si<s_j(0)$. (There will also be lower limits on the ranges of $\si$, determined by boundary conditions at the other ends of the strings, but for the moment we assume that they are far enough away to be irrelevant.)  In the subsequent motion the amplitudes of the inward-moving waves at the vertex, namely $\bb'_j(s_j-t)$, will thus be known, but the amplitudes of the outgoing waves $\ba'_j(s_j+t)$ will not.  They can be found from the various junction conditions as follows.

First, differentiating (\ref{abX}), we find [at $\si=s_j(t)$]
 \beq (1+\dot s_j)\ba'_j-(1-\dot s_j)\bb'_j=2\bXd. \label{abXdot}\eeq
Substituting for $\ba'_j$ from this equation into (\ref{sumab}), we get
 \beq \sum_j\mu_j(1-\dot s_j)\bb'_j=-(\mu_1+\mu_2+\mu_3)\bXd.
 \label{sumbX} \eeq
Eliminating $\bXd$ from (\ref{abXdot}) and (\ref{sumbX}), each $\ba'_j$ can be expressed as a linear combination of the $\bb'_j$. We still have to determine the unknown values of $\dot s_j$, but this can be done by imposing the gauge conditions $\ba'_j{}^2=1$. The result is a function of the string tensions $\mu_j$, as well as the scalar products
 \beq c_{ij}=\bb'_i(s_i-t)\cd\bb'_j(s_j-t) =c_{ji}. \label{c}\eeq
It turns out to be simplest to solve for the three unknowns $1-\dot s_j$.  To express the result concisely, we define three combinations $M_j$ of the string tensions by
 \beq M_1=\mu_1^2-(\mu_2-\mu_3)^2, \eeq
and two similar equations for $M_2$ and $M_3$.  Then we find
 \bea &&\frac{\mu_1(1-\dot s_1)}{\mu_1+\mu_2+\mu_3}\nonumber\\
 &=&\frac{M_1(1-c_{23})}{M_1(1-c_{23})+M_2(1-c_{31})+M_3(1-c_{12})},
 \label{sdot} \eea
together with two similar equations obtained by cyclic permutation.

There is one important immediate corollary, which follows from the obvious restriction $\dot s_j\le 1$.  This implies that each $M_j\ge 0$.  In other words, the three string tensions must satisfy the triangle inequalities; if one tension exceeds the sum of the other two, no three-string junction is possible.  This is of course obvious for the case of a static equilibrium configuration.

Note that summing the three equations (\ref{sdot}) yields the relation
 \beq \mu_1\dot s_1+\mu_2\dot s_2+\mu_3\dot s_3=0.\label{sumsdot} \eeq
This is an expression of energy conservation: the rate of creation of new string must balance the disappearance of old.

The equations (\ref{sdot}) serve to determine the values of $s_j(t)$.  Note that these are differential equations for $s_j$ rather than an explicit solution, because in the light of the definition (\ref{c}) the values of $s_j(t)$ also appear on the right-hand side.  So in general a numerical solution may be needed.  Once we have found the $s_j(t)$, we can at once write down the values of $\ba'_j(s_j+t)$ from (\ref{abXdot}) and (\ref{sumbX}), and then integrate to find $\ba_j$.

Of course, this process can only proceed so long as the relevant values of $\bb'_j$ are within the range determined by the initial conditions.  Eventually the effects of other junctions will come into play, and the values of $\bb'_j$ will be ones determined by earlier dynamics at these other junctions, not by the initial conditions.  Nevertheless, an iterative solution of the dynamical equations for all the junctions together is in principle feasible.

As an example, it is easy to generalize the familiar collapsing circular loop solution to a configuration comprising three semicircular arcs, namely
 \beq \bx_j(\si,t)=\cos t(\cos\si\cos\theta_j,\cos\si\sin\theta_j,\sin\si),
 \label{3arcs} \eeq
where $|\si|\le\pi/2$, and the angles $\theta_j$ are chosen to satisfy the equilibrium conditions
 \beq \sum_j\mu_j e^{i\theta_j}=0. \label{wsum} \eeq
This is always possible provided the $\mu_j$ satisfy the triangle
inequalities.  Clearly, here we can take
$\ba_j(\si)=\bb_j(\si)=\bx_j(\si,0)$.  From (\ref{c}) and (\ref{wsum})
it is then straightforward to verify that, when each $s_j=\pi/2$,
(\ref{sdot}) implies that all $\dot s_j=0$, so (\ref{3arcs}) is a self-consistent solution.  The loops remain semicircular and shrink to  a point at time $t=\pi/2$.

We now turn to our central problem: what happens when two strings that can exchange partners, becoming linked by a third string, meet?   In general, this can happen in two different ways, and it is not obvious which one is chosen, or indeed whether they do exchange partners at all.

Consider two straight strings approaching one another along the $z$-axis.  For simplicity, we discuss in this paper the case of equal tension, $\mu_1=\mu_2$, although similar results hold more generally (as we shall describe in a future publication).  For $t<0$, we take
 \beq \bx_{1,2}(\si,t)=(-\ga^{-1}\si\cos\al,
 \mp\ga^{-1}\si\sin\al,\pm vt).\label{initial} \eeq
Here $v$ is the string velocity and $\ga^{-1}=\sqrt{1-v^2}$.  Thus
 \bea \ba'_{1,2}&=&(-\ga^{-1}\cos\al,\mp\ga^{-1}\sin\al,\pm v),
 \nonumber\\
 \bb'_{1,2}&=&(-\ga^{-1}\cos\al,\mp\ga^{-1}\sin\al,\mp v). \eea
(The sign of $\si$ is chosen to match our earlier conventions.)

After the strings cross, we expect that they will be joined by a third string, which by symmetry must lie either along the $x$-axis or along the $y$-axis.  \emph{A priori}, we might guess that if the angle $\al$ is small, the connecting string would be in the $x$ direction, while if it is closer to $\pi/2$, it would choose the $y$ direction.  We shall see that this is partly correct.

\begin{figure}
\includegraphics{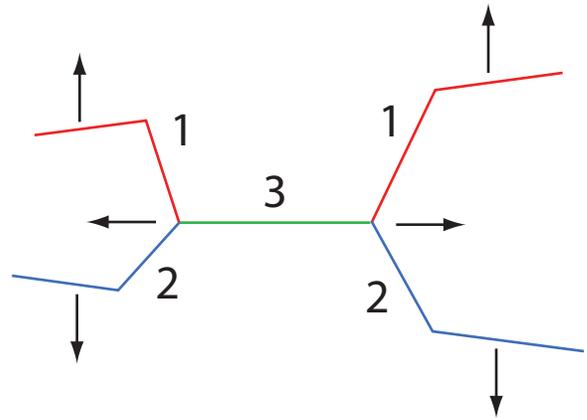}
\caption{Two strings joined by a third after intercommuting.}
\end{figure}

To be specific, let us suppose that the connecting string (labeled 3) is along the $x$-axis.  (See Fig.~1.)  Then clearly, in this region,
 \beq \bx_3(\si,t)=(\si,0,0),\qquad
 \ba'_3(\si)=\bb'_3(\si)=(1,0,0). \eeq
There will be two junctions at the ends of this string, symmetrically placed either side of the origin.  By the symmetry of the problem it is sufficient to consider one of them; we shall take the one on the positive $x$-axis.
Clearly, using (\ref{sumsdot}), $s_1=s_2=-(\mu_3/2\mu_1)s_3$.  The vertex position is $\bX(t)=(s_3(t),0,0)$.  It moves along the $x$-axis with uniform velocity $\dot s_3$.

To apply the previous method, we first evaluate the $c_{ij}$ of
(\ref{c})
and then substitute into (\ref{sdot}).  This yields
 \beq \dot s_3=\frac{2\mu_1\ga^{-1}\cos\al-\mu_3}
 {2\mu_1-\mu_3\ga^{-1}\cos\al},\qquad
 \dot s_1=\dot s_2=-\frac{\mu_3}{2\mu_1}\dot s_3.
 \label{sdot3} \eeq
On the strings 1 and 2, for $t>0$ there are kinks at $\si=t$, at the positions $t(\ga^{-1}\cos\al,\pm\ga^{-1}\sin\al,\pm v)$, beyond which the expressions (\ref{initial}) still apply.  The vertex $\bX$ is joined to these kinks by new straight segments. From (\ref{abXdot}) one can find values of $-\bb'_{1,2}$ and $\ba'_3$ representing the outgoing waves, and verify that they are consistent.

This result has interesting implications.  It is clear that the solution only makes sense if $\dot s_3>0$: the connecting string 3 cannot get \emph{shorter}.  Thus we require
 \beq \al<\arccos\left(\frac{\mu_3\ga}{2\mu_1}\right)\qquad
 (x\mathrm{-axis}). \eeq
This is in line with our expectation that the connecting string would form on the $x$-axis for small $\al$.  For a string along the $y$-axis one would require
 \beq \al>\arcsin\left(\frac{\mu_3\ga}{2\mu_1}\right)\qquad
 (y\mathrm{-axis}). \eeq
As we have already noted, no junction is possible if $\mu_3>2\mu_1$. Moreover, for any mass ratio there is a limiting velocity above which a junction cannot form; we require
 \beq \ga<\frac{2\mu_1}{\mu_3}.\label{limv1} \eeq
For example, if the tensions are all equal, no junction can form unless $v<\sqrt3/2$.  Strings approaching each other at very large velocity \emph{cannot} exchange partners.  Abelian strings will simply pass through one another.  For non-abelian strings, if this is topologically forbidden, they may become joined by a new string along the $z$-axis, but \emph{without} exchanging partners.  It turns out that this process is kinematically allowed provided that
 \beq v > \frac{\mu_3}{2\mu_1}, \qquad
 (z\mathrm{-axis}),\label{limv2} \eeq
independent of $\alpha$.  Note that if $\mu_3$ is large there may be a range of velocities for which neither inequality (\ref{limv1}) nor (\ref{limv2}) can be satisfied.  In that case, the strings must be locked into an X configuration, unable to separate in any direction.  This may have important implications for the evolution of a cosmological network of such strings.

It is worth noting that there are cases in which two or more of these configurations are possible.  In such cases the choice may be random.

The discussion can be extended to the collision of two strings of different tension, $\mu_1\ne\mu_2$, with similar results, although in general we have not found an analytic formula for the limiting velocity.  We shall discuss the general case in a later paper.

We have shown that there are important kinematic constraints on the possibility of intercommuting of strings that form junctions. If the relative velocity with which they meet is too large, no exchange can take place and the strings will merely pass through one another (or, for non-abelian strings, become joined by a string in the direction of the relative velocity or form a linked X configuration).  This restriction, and the difference in behavior between non-abelian strings and junction-forming abelian strings, may be of considerable importance in studies of the evolution of a cosmological string network.  Of course, even if the kinematic constraints are satisfied, there is no guarantee that intercommuting will occur. This dynamical problem has been discussed by several authors \cite{Jackson:2004zg,Hanany:2005bc} with the conclusion that the intercommuting probability is frequently much less than one.  The effect of reducing the intercommuting probability on the evolution of a network of standard cosmic strings (with no junctions) has been considered in \cite{Jones:2003da,Damour:2004kw,Sakellariadou:2004wq,Avgoustidis:2005nv}.

\begin{acknowledgments}
We are grateful to Tanmay Vachaspati for a valuable comment.
The work reported here was assisted by the ESF COSLAB Programme.
\end{acknowledgments}


\begin{thebibliography}{99}

\bibitem{Sazhin:2003cp}
M.~Sazhin {\it et al.},
Mon.\ Not.\ Roy.\ Astron.\ Soc.\  {\bf 343}, 353 (2003)
[arXiv:astro-ph/0302547].

\bibitem{Sazhin:2004fv}
M.~V.~Sazhin, O.~S.~Khovanskaya, M.~Capaccioli, G.~Longo, J.~M.~Alcala, R.~Silvotti and M.~V.~Pavlov,
arXiv:astro-ph/0406516.

\bibitem{Polchinski:2004hb}
J.~Polchinski,
AIP Conf.\ Proc.\  {\bf 743}, 331 (2005)
[Int.\ J.\ Mod.\ Phys.\ A {\bf 20}, 3413 (2005)]
[arXiv:hep-th/0410082].

\bibitem{Kibble:2004hq}
T.~W.~B.~Kibble,
Lecture at COSLAB 2004, Ambleside, 10--17 September 2004 (unpublished),
arXiv:astro-ph/0410073.

\bibitem{Davis:2005dd}
A.~C.~Davis and T.~W.~B.~Kibble,
Contemp.\ Phys.\  {\bf 46}, 313 (2005)
[arXiv:hep-th/0505050].

\bibitem{Hindmarsh:1994re}
M.~B.~Hindmarsh and T.~W.~B.~Kibble,
Rept.\ Prog.\ Phys.\  {\bf 58}, 477 (1995)
[arXiv:hep-ph/9411342].

\bibitem{Vilenkin:1994}
A.~Vilenkin and E.~P.~S.~Shellard,
\emph{Cosmic Strings and other Topological Defects}
(Cambridge University Press, Cambridge, 1994).

\bibitem{Sazhin:2006fe}
M.~V.~Sazhin {\it et al.},
arXiv:astro-ph/0601494.

\bibitem{Copeland:2003bj}
E.~J.~Copeland, R.~C.~Myers and J.~Polchinski,
JHEP {\bf 0406}, 013 (2004)
[arXiv:hep-th/0312067].

\bibitem{Alford:1992yx}
M.~G.~Alford {\it et al.},
Nucl.\ Phys.\ B {\bf 384}, 251 (1992)
[arXiv:hep-th/9112038].

\bibitem{Hashimoto:2005hi}
K.~Hashimoto and D.~Tong,
JCAP {\bf 0509}, 004 (2005)
[arXiv:hep-th/0506022].

\bibitem{Vachaspati:1986cc}
T.~Vachaspati and A.~Vilenkin,
Phys.\ Rev.\ D {\bf 35}, 1131 (1987).

\bibitem{Spergel:1996ai}
D.~Spergel and U.~L.~Pen,
Astrophys.\ J.\  {\bf 491}, L67 (1997)
[arXiv:astro-ph/9611198].

\bibitem{Tye:2005fn}
S.~H.~Henry Tye, I.~Wasserman and M.~Wyman,
Phys.\ Rev.\ D {\bf 71}, 103508 (2005);
\emph{ibid.}\ {\bf 71}, 129906(E) (2005)
[arXiv:astro-ph/0503506].

\bibitem{Copeland:2005cy}
E.~J.~Copeland and P.~M.~Saffin,
JHEP {\bf 0511}, 023 (2005)
[arXiv:hep-th/0505110].

\bibitem{Austin:1994dz}
D.~Austin, E.~J.~Copeland and T.~W.~B.~Kibble,
Phys.\ Rev.\ D {\bf 51}, R2499 (1995)
[arXiv:hep-ph/9406379].

\bibitem{'tHooft:2004he}
G.~'t Hooft,
arXiv:hep-th/0408148.

\bibitem{Kibble:1982cb}
T.~W.~B.~Kibble and N.~Turok,
Phys.\ Lett.\ B {\bf 116}, 141 (1982).

\bibitem{Burden:1985md}
C.~J.~Burden,
Phys.\ Lett.\ B {\bf 164}, 277 (1985).

\bibitem{Shellard:1987bv}
E.~P.~S.~Shellard,
Nucl.\ Phys.\ B {\bf 283}, 624 (1987).

\bibitem{Polchinski:1988cn}
J.~Polchinski,
Phys.\ Lett.\ B {\bf 209}, 252 (1988).

\bibitem{Copeland:1986ng}
E.~J.~Copeland and N.~Turok,
FERMILAB-PUB-86-127-A (1986).

\bibitem{Shellard:1988zx}
E.~P.~S.~Shellard and P.~J.~Ruback,
Phys.\ Lett.\ B {\bf 209}, 262 (1988).

\bibitem{Matzner:1988}
R.~A.~Matzner,
Computers in Physics, Sep/Oct, 51 (1988).

\bibitem{Jackson:2004zg}
M.~G.~Jackson, N.~T.~Jones and J.~Polchinski,
JHEP {\bf 0510}, 013 (2005)
[arXiv:hep-th/0405229].

\bibitem{Hanany:2005bc}
A.~Hanany and K.~Hashimoto,
JHEP {\bf 0506}, 021 (2005)
[arXiv:hep-th/0501031].

\bibitem{Jones:2003da}
N.~T.~Jones, H.~Stoica and S.~H.~Henry Tye,
Phys.\ Lett.\ B {\bf 563}, 6 (2003)
[arXiv:hep-th/0303269].

\bibitem{Damour:2004kw}
T.~Damour and A.~Vilenkin,
Phys.\ Rev.\ D {\bf 71}, 063510 (2005)
[arXiv:hep-th/0410222].

\bibitem{Sakellariadou:2004wq}
M.~Sakellariadou,
JCAP {\bf 0504}, 003 (2005)
[arXiv:hep-th/0410234].

\bibitem{Avgoustidis:2005nv}
A.~Avgoustidis and E.~P.~S.~Shellard,
arXiv:astro-ph/0512582.


\end{thebibliography}
\end{document}